\documentclass[preprint2]{aastex}
\usepackage{epsfig}
\usepackage{color}

\newcommand   {\about} {\mbox{$\sim$}}
\newcommand   {\mic}   {\mbox{$\mu$m}}

\newcommand   {\hh}    {\mbox{H$_2$}}

\newcommand   {\hcop}  {\mbox{HCO$^+$}}
\newcommand   {\nndp}  {\mbox{N$_2$D$^+$}}
\newcommand   {\nnhp}  {\mbox{N$_2$H$^+$}}

\newcommand   {\hhdp}  {\mbox{H$_2$D$^+$}}

\newcommand   {\nhhh}  {\mbox{NH$_3$}}
\newcommand   {\nddd}  {\mbox{ND$_3$}}
\newcommand   {\nddh}  {\mbox{ND$_2$H}}
\newcommand   {\nhhd}  {\mbox{NH$_2$D}}

\newcommand   {\hthcop}{\mbox{H$^{13}$CO$^+$}}

\newcommand   {\dcop}  {\mbox{DCO$^+$}}
\newcommand   {\hho}   {\mbox{H$_2$O}}

\renewcommand {\deg}   {\mbox{$^\circ$}}
\newcommand   {\arcm}  {\mbox{$^\prime$}}
\newcommand   {\arcs}  {\mbox{$^{\prime\prime}$}}
\newcommand   {\pccm}  {\mbox{cm$^{-3}$}}
\newcommand   {\pscm}  {\mbox{cm$^{-2}$}}
\newcommand   {\kms}   {\mbox{km\,s$^{-1}$}}

\renewcommand {\ga}    {\mbox{\rlap{\hbox{\lower5pt\hbox{$\sim$}}}\hbox{$>$}}}
\renewcommand {\la}    {\mbox{\rlap{\hbox{\lower5pt\hbox{$\sim$}}}\hbox{$<$}}}
\newcommand {\arcsper} {\mbox{\rlap{\hbox{\hbox{.}}}\hbox{$^{\prime\prime}$}}}
\newcommand   {\hr}    {\mbox{$^{\rm h}$}}
\newcommand   {\mt}    {\mbox{$^{\rm m}$}}
\newcommand {\secper}  {\mbox{\rlap{\hbox{\hbox{.}}}\hbox{$^{\rm s}$}}}
\newcommand  {\solar}  {\mbox{$_{\odot}$}}

\received{2016 March 14}
\accepted{2016 May 3}

\slugcomment{Accepted for publication in the Astrophysical Journal
  2016 May 3}

\shortauthors{Lis et al.}

\shorttitle{Prestellar core in L1689N}

\begin{document}

\title{Star Formation and Feedback: A Molecular Outflow--Prestellar Core
  Interaction in L1689N}

\author{D.~C.~Lis$^{1,2}$, H.~A.~Wootten$^3$, M.~Gerin$^4$,
  L.~Pagani$^1$, E.~Roueff$^5$, F.F.S~van~der~Tak$^{6,7}$, C.~Vastel$^8$, and
  C.M.~Walmsley$^{9,10}$ }

\altaffiltext{1}{LERMA, Observatoire de Paris, PSL Research
  University, CNRS, Sorbonne Universit\'{e}s, UPMC Univ. Paris 06, F-75014
  Paris, France; darek.lis@obspm.fr} 

\altaffiltext{2}{California Institute of Technology, Cahill Center for
  Astronomy and Astrophysics 301-17, Pasadena, CA~91125, USA}

\altaffiltext{3}{National Radio Astronomy Observatory,
  Charlottesville, VA 22903, USA; awootten@nrao.edu}

\altaffiltext{4}{LERMA, Observatoire de Paris, PSL Research
  University, CNRS, Sorbonne Universit\'{e}s, UPMC Univ. Paris 06, \'{E}cole
  normale sup\'{e}rieure, F-75005, Paris, France;
  maryvonne.gerin@lra.ens.fr} 

\altaffiltext{5}{LERMA, Observatoire de Paris, PSL Research
  University, CNRS, Sorbonne Universit\'{e}s, UPMC Univ. Paris 06,
  F-92190 Meudon, France}

\altaffiltext{6} {SRON Netherlands Institute for Space Research,
  Landleven 12, 9747 AD Groningen, The Netherlands}

\altaffiltext{7}{Kapteyn Astronomical Institute, University of
  Groningen, The Netherlands} 

\altaffiltext{8}{Institut de Recherche en Astrophysique et Planetologie,
F-31028 Toulouse Cedex 4, France}

\altaffiltext{9}{INAF -– Osservatorio astrofisico di Arcetri, Largo E.
  Fermi 5, I-50125 Firenze, Italy} 

\altaffiltext{10}{Dublin Institute of Advanced Studies, Fitzwilliam
  Place 31, Dublin 2, Ireland}

\begin{abstract}
  We present \emph{Herschel}\footnote{\emph{Herschel} is an ESA space
    observatory with science instruments provided by European-led
    Principal Investigator consortia and with important participation
    from NASA.}, ALMA Compact Array (ACA), and Caltech Submillimeter
  Observatory (CSO) observations of the prestellar core in L1689N,
  which has been suggested to be interacting with a molecular outflow
  driven by the nearby solar type protostar IRAS~16293-2422. This
  source is characterized by some of the highest deuteration levels
  seen in the interstellar medium. The change in the \nhhd\ line
  velocity and width across the core provides clear evidence of an
  interaction with the outflow, traced by the high-velocity water
  emission. Quiescent, cold gas, characterized by narrow line widths
  is seen in the NE part of the core, while broader, more disturbed
  line profiles are seen in the W/SW part. Strong \nndp\ and \nddd\
  emission is detected with the ACA, extending S/SW from the peak of
  the single-dish \nhhd\ emission. The ACA data also reveal the
  presence a compact dust continuum source, with a mean size of \about
  1100 au, a central density of $(1-2)\times 10^7$~\pccm, and a mass
  of 0.2--0.4~M\solar. The dust emission peak is displaced \about
  5\arcs\ to the south with respect to the \nndp\ and \nddd\ emission,
  as well as the single-dish dust continuum peak, suggesting that the
  northern, quiescent part of the core is characterized by spatially
  extended continuum emission, which is resolved out by the
  interferometer. We see no clear evidence of fragmentation in this
  quiescent part of the core, which could lead to a second generation
  of star formation, although a weak dust continuum source is detected
  in this region in the ACA data.
\end{abstract}


\section{Introduction}

Low-mass star formation is known to occur exclusively in the shielded
interiors of molecular cloud cores, when gravity wrests control from
supporting thermal, magnetic and turbulent pressures and collapse
ensues. Over the past 20 years, much effort has been focused on
understanding the subsequent stages of star formation, when a central
object and a surrounding circumstellar disk are formed and evolve
toward a newly formed star and, potentially, its associated planetary
system. However, the initiation of this process is among the least
understood steps of star formation. Yet, it is key to understanding
some of its most fundamental aspects, such as the initial mass
function, the binarity fraction and its dependence on stellar mass,
and the star formation efficiency. 

Generally, starless cores are thought to represent this earliest stage
of star formation. In one view, magnetically supported clouds develop
magnetically critical cores over a long timescale (a few Myr) in
quasi-static fashion through the process of ambipolar diffusion
\citep{shu87, mousch91}. Alternatively, turbulence continuously shreds
and twists molecular clouds, forming denser structures--through
shocks--that collapse, fragment or otherwise disappear. In this view,
starless cores are highly dynamical structures that evolve in a few
sound-crossing times, i.e. 10 times faster than the ambipolar
diffusion time scale (see \citealt{balles07} and references therein).
These different scenarios for the formation and evolution of starless
cores result in very different 3-D morphological structures and unique
dynamic signatures. The environment in which a core forms is also
critical to its subsequent evolution. Dense cores in regions of
cluster-forming clouds tend to have higher masses and column densities
than isolated prestellar cores. The molecular cloud cores in Taurus
and Ophiuchus provide a case in point (e.g., \citealt{ward07} and
references therein).

Much of our insight into the structure of starless cloud cores comes
from observations of the millimeter dust emission \citep{bergin07}.
\emph{Herschel} observations using PACS and SPIRE have shown that
cores form preferentially along dense filaments \citep{andre10,
  molinari10}. The \emph{Herschel} Gould belt survey in particular has
discovered a large population of nearby prestellar cores, allowing
accurate determination of the core mass function (e.g.,
\citealt{konyves10, andre14}, but see the recent results of
\citealt{pagani15}), which can be compared with model predictions for
the IMF produced by the collapse of a turbulent cloud (e.g.,
\citealt{hennebelle08, hennebelle09}). High angular resolution
interferometric continuum studies give insights into the formation of
multiple systems \citep{maury10}. However, continuum observations are
likely to yield only a partial picture of the cloud structure since
coagulation of dust is a key process at the high densities of inner
starless cores and this will change the grain opacity coefficient and
effectively “hide” much of the mass of the dust from view
\citep{steinacker10, pagani10, lefevre14, lefevre16}. Moreover, dust
studies do not provide direct insight into the dynamics of these cores
nor into their chemistry.

Likewise, molecular observations are also known, in general, to
provide a biased view of starless cores. This reflects the
condensation of species onto ice mantles at high densities. There are
two exceptions to this general rule. First, for reasons that are not
fully understood, nitrogen-bearing species, in particular ammonia, do
not seem to participate in this freeze-out \citep{tafalla02,
  crapsi07}. Deuterated species form the second exception (see the
recent review by \citealt{ceccarelli14}). High deuteration of
gas-phase species is, in effect, a result of the freeze-out, and
coincidental disappearance of ortho-\hh\ from the gas phase, without
which no deuteration would happen \citep{pagani92, flower06,
  pagani09}. This drives up the gas-phase abundance and fractionation
of H$_3^+$. The high fractionation of this species is then passed on
to the few molecules remaining in the gas phase, through ion-molecule
chemistry \citep{roberts00}. These chemical aspects can be used to our
advantage to study prestellar cores. We note that some deuterated
species, such as formaldehyde or water, are formed and deuterated
primarily through grain-surface chemistry in icy mantles
\citep{ceccarelli14}. These molecules can then be thermally or
photo-desorbed in active environments, such as molecular hot cores,
outflows, or shocks. Of the two key deuterated species studied here,
\nndp\ is believed to be formed in cold, dense, CO-depleted regions
through gas-phase chemistry, driven by H$_3^+$ and its deuterated
isotopologues. Ammonia deuteration is also straightforward to
understand in the framework of gas-phase processes \citep{roueff05}.
However, recent laboratory experiments \citep{fedoseev15} suggest that
ammonia and its deuterated isotopologues may also be formed thorough
successive additions of hydrogen and deuterium atoms to nitrogen atoms
in CO-rich interstellar ice analogues, which are efficient at
temperatures below 15~K.

High spectral resolution investigations of the velocity field in
central regions of cold, dense cores are invaluable for disentangling
rotation and collapse \citep{belloche02}, providing excellent
constraints on fragmentation and formation of disks, which are
critically dependent on the magnetic field \citep{hennebelle07a,
  hennebelle07b}. This requires, however, that suitable molecular
tracers be chosen. The fundamental, submillimeter lines of ammonia
isotopologues have critical densities of order $10^6$~\pccm\ and, with
simple hyperfine patterns, are excellent tracers of dense gas at the
onset of star formation \citep{lis02b, lis06}, providing information
complementary to H$_2$D$^+$ \citep{vastel06, caselli08, vandertak05,
  vastel12}. The interferometric study of \cite{crapsi07} has shown a
very high NH$_2$D/NH$_3$ fractionation ratio in the central region of
L1544, $\about 0.5\pm 0.2$, which demonstrates the persistence of
deuterated ammonia molecules at very high densities (\about
$2 \times 10^6$ \pccm) and low temperatures (\about 10~K). Very high
deuteration ratios have also been reported in other molecules (e.g.
\nndp/\nnhp$= 0.7\pm 0.2$ in L183; \citealt{pagani07}; see
the recent review of \citealt{ceccarelli14}). Nevertheless,
lines of deuterated isotopologues of ammonia, with their simple
hyperfine patterns, fulfill all requirements for being optimum tracers
of cold, dense cores.

\emph{Herschel} has enabled for the first time systematic studies of
the fundamental submillimeter lines of NH$_3$ and NH$_2$D in
prestellar cores. The long-term goals of this study are threefold: (a)
to firmly establish the utility of ammonia isotopologues as tracers of
dense, cold gas at the onset of star formation, as postulated by
\cite{lis06, lis08}; (b) to determine the morphology and velocity
structure of an example prestellar core, the processes that drive the
evolution in this earliest stage of star formation and that control
such important aspects as core fragmentation, the resulting initial
mass function and binarity fraction of protostars; (c) to improve our
understanding of the chemical and physical processes that control
gas-grain interaction, freeze out, and mantle ejection in dense cloud
cores, including when/why species freeze out, and of deuterium
fractionation. These processes are of key interest for the chemistry
of regions of star formation and, hence, the supply of volatiles in
planet-forming environments including likely the early Solar Nebula.

In this paper, we present extensive single-dish and interferometric
observations of L1689N, a nearby (120~pc; \citealt{loinard08}) dark
cloud located in the $\rho$~Ophiuchi complex, which harbors one of the
best-studied solar type protostars IRAS~16293-2422 (hereafter
IRAS~16293). The nearby protostellar core, sometimes referred to as
I16293E, has been previously suggested to be interacting with the blue
lobe of one of the outflows driven by IRAS~16293 \citep{wootten87,
  lis02a, stark04}. This region is thus an excellent laboratory to
study the effects of stellar feedback on the physics and chemistry of
the surrounding interstellar medium (ISM). In addition to
\emph{Herschel} observations, we have obtained interferometric
observations of the 970~\mic\ dust continuum, \nddd, and N$_2$D$^+$
line emission using the ALMA Compact Array, as well as maps of
several deuterated molecular tracers, using the Caltech Submillimeter
Observatory. The observational results are presented in
Section~\ref{sec:results}, followed by the discussion in
Section~\ref{sec:discussion}, and a summary in
Section~\ref{sec:summary}.

\section{Observations}

\begin{table*}[!t]
\small
\begin{center}
\caption{\emph{Herschel}, ACA, and CSO Observations of L1689N.}
\label{tab:spectro}
\vskip 2mm 
\begin{tabular}{lccccc}
\hline\hline
Species & Frequency & Transition &  E$\rm_{u}$ \\ 
             & (GHz)        &                  & (K) \\ 
  \hline
\multicolumn{4}{c}{\textit{Herschel}}\\
  \hline
p-NH$_2$D              & 494.45455 
                    & $1_{1,0}-0_{0,0}$ (0a--0s) & 23.7\\ 
o-H$_2$O                & 556.93599 
                    & $1_{1,0}-1_{0,1}$               & 61.0\\ 
o-NH$_3$                & 572.49816 
                    & $1_0-0_0$ (0s--0a)        & 27.5\\ 
SO                            & 558.08764 
                    & $12_{13}-11_{12}$            & 194.4\\ 
SO                            & 559.31952 
                    & $13_{13}-12_{12}$            & 201.1\\ 
SO                            & 560.17865 
                    & $14_{13}-13_{12}$            & 192.7\\ 
  \hline
\multicolumn{4}{c}{\textit{ALMA Compact Array}}\\
  \hline
N$_2$D$^+$   & 308.42227 
                    &  4--3                                & 37.0\\ 
CH$_3$OH      & 309.29040 
                    & $5_{1,4} - 5_{0,5}$ A -- +   & 49.7\\ 
SO                   & 309.50244 
                    & $2_2-2_1$                        & 19.3\\ 
ND$_3$           & 309.90949 
                    & $1_0-0_0$ (0a - 0s)         & 14.9\\ 
Continuum      & 309.963    &  &   \\
Continuum         & 353.949    &  &   \\
\hline
\multicolumn{4}{c}{\textit{Caltech Submillimeter Observatory}}\\
\hline
\dcop\      &  216.11258 & 3--2                 & 10.4 \\
DCN          &  217.23854 & 3--2                 & 10.4 \\
DNC          &  228.91048 & 3--2                 & 11.0 \\
\nndp\      &  231.32183 & 3--2                 & 11.1 \\
HCN          &  265.88643 & 3--2                 & 12.8 \\
\hcop\      &  267.55763 & 3--2                 & 12.8 \\
HNC          &  271.98114 & 3--2                 & 13.1 \\
\nnhp\      &  279.51175 & 3--2                 & 13.4 \\
\hline
\end{tabular}
\end{center}
\end{table*}

The observations of the fundamental rotational transitions of \nhhd,
\nhhh, and o-\hho\ in L1689N presented here\footnote{\emph{Herschel}
  OBSIDs: 1342238591--38592, 38650, 50470, 51430; open time program
  \emph{Ammonia as a Tracer of the Earliest Stages of Star Formation},
  OT1\_dlis\_2, OT2\_dlis\_3.} were carried out in 2012
February--September, using the Band 1 receivers of the Heterodyne
Instrument for the Far-Infrared (HIFI; \citealt{degraauw10}) on the
\emph{Herschel} Space Observatory \citep{pilbratt10}. Both the WBS and
HRS spectrometers were used in parallel. The WBS provides full
coverage of the 4~GHz intermediate frequency band in the upper and
lower sidebands with a 1.1~MHz resolution (0.59~\kms\ at 557~GHz),
while the HRS provides higher spectral resolution spectra of selected
lines of interest (spectral resolution of 0.12~MHz or 0.07~\kms\ at
492~GHz). For mapping observations (\about $180 \times 190$\arcs\ for
water, $70 \times 90$\arcs\ for \nhhd), the HIFI OTF observing mode
was used, while the frequency-switching mode was used for deeper,
pointed observations. The data have been processed through the
standard HIFI data reduction pipeline using HIPE version 10.0
\citep{ott10} and the resulting spectra were subsequently reduced
using the GILDAS CLASS\footnote{http://www.iram.fr/IRAMFR/GILDAS}
software package. The FWHM HIFI beam size is \about 44\arcs\ at 492
GHz and 38\arcs\ at 557~GHz and the main beam efficiency is 76\%
\citep{roelfsema12}. The frequencies and quantum numbers of the
transitions observed are listed in Table~\ref{tab:spectro}.

ALMA observations were obtained within two frequency settings (science
goals), one near 310~GHz and a second near 340~GHz. The spectral lines
discussed in the present manuscript correspond to the 310~GHz
frequency setting and are summarized in Table~\ref{tab:spectro}. A ten
pointing mosaic covering a \about $65 \times 80$\arcs\ region was
observed on seven occasions during 2014 January--April, using eleven
7-m antennas of the Morita Array component of ALMA in Cycle 1. At each
point of the mosaic, 15 minutes of data was collected for a
sensitivity of 0.41~Jy/beam in a 0.1~\kms\ channel. Four spectral
windows were observed centered near 309 GHz, with varying spectral
resolution. The N$_2$D$^+$ $J=4-3$ line was in a 0.0625~GHz width
spectral window, centered near 308.449 GHz (velocity resolution of
0.045~\kms), and the ND$_3$ line lay in another 0.0625~GHz width
spectral window, centered near 309.94 GHz. The 970~\mic\ dust
continuum flux was extracted by averaging channels within two bands of
1.992~GHz width, centered near 308.053 and 309.873~GHz. CH$_3$OH and
H$_3$O$^+$ lines in the bands were flagged before averaging to avoid
possible contamination. In the continuum image, the peak flux is
76~mJy/beam and the integrated flux is 840~mJy. The rms in the image
is 1.1~mJy/beam (an excellent SNR of 70) and the synthesized beam size
is $5.04 \times 3.16$\arcs, at a position angle of 80\deg.

\begin{figure*}[!t]
\centering
\includegraphics[width=0.7\textwidth,angle=0]{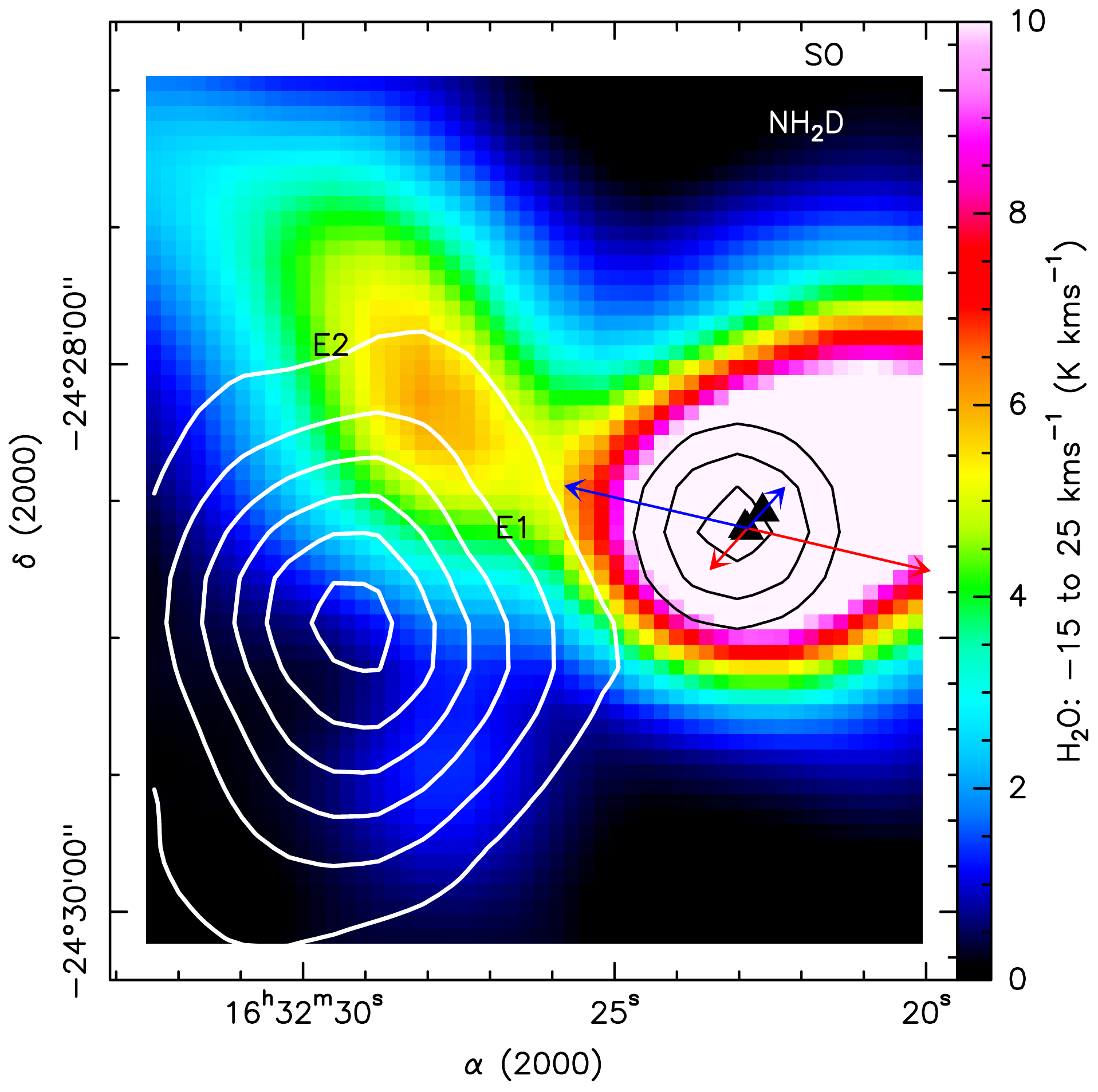}
\caption{Color image of the integrated line intensity of the
  fundamental rotational transition of o-\hho, between -15 and
  25~\kms, with overlaid white contours of the integrated intensity of
  the fundamental rotational transition of \nhhd, between 1.7 and
  5.5~\kms\ (including the 3 hyperfine components). The \nhhd\
  emission shows the location of the prestellar core, while the
  location of the solar type protostar IRAS~16293 driving the
  molecular outflow is shown by the black contours of the excited SO
  emission (average of the three transitions listed in
  Table~\ref{tab:spectro}). Contour levels for \nhhd\ are from 20\% to
  95\% of the peak (0.65~K\,\kms), with a step of 15\%. For SO, contour
  levels are 50\%, 70\% and 90\% of the peak (2.25~K\,\kms). Black
  triangles mark the locations of IRAS~16293 A and B (lower and upper,
  respectively). The red and blue arrows mark the directions of the
  compact CO outflows observed with the SMA \citep{girart14}, which
  have opposite polarity to the large-scale outflows seen in earlier
  single-dish studies (e.g. \citealt{stark04}). E1 and E2 mark the
  locations of the two SiO clumps of \cite{hirano01}.}
\label{fig:morphology}
\end{figure*}

\begin{figure*}[!ht]
\centering
\includegraphics[width=0.95\textwidth,angle=0]{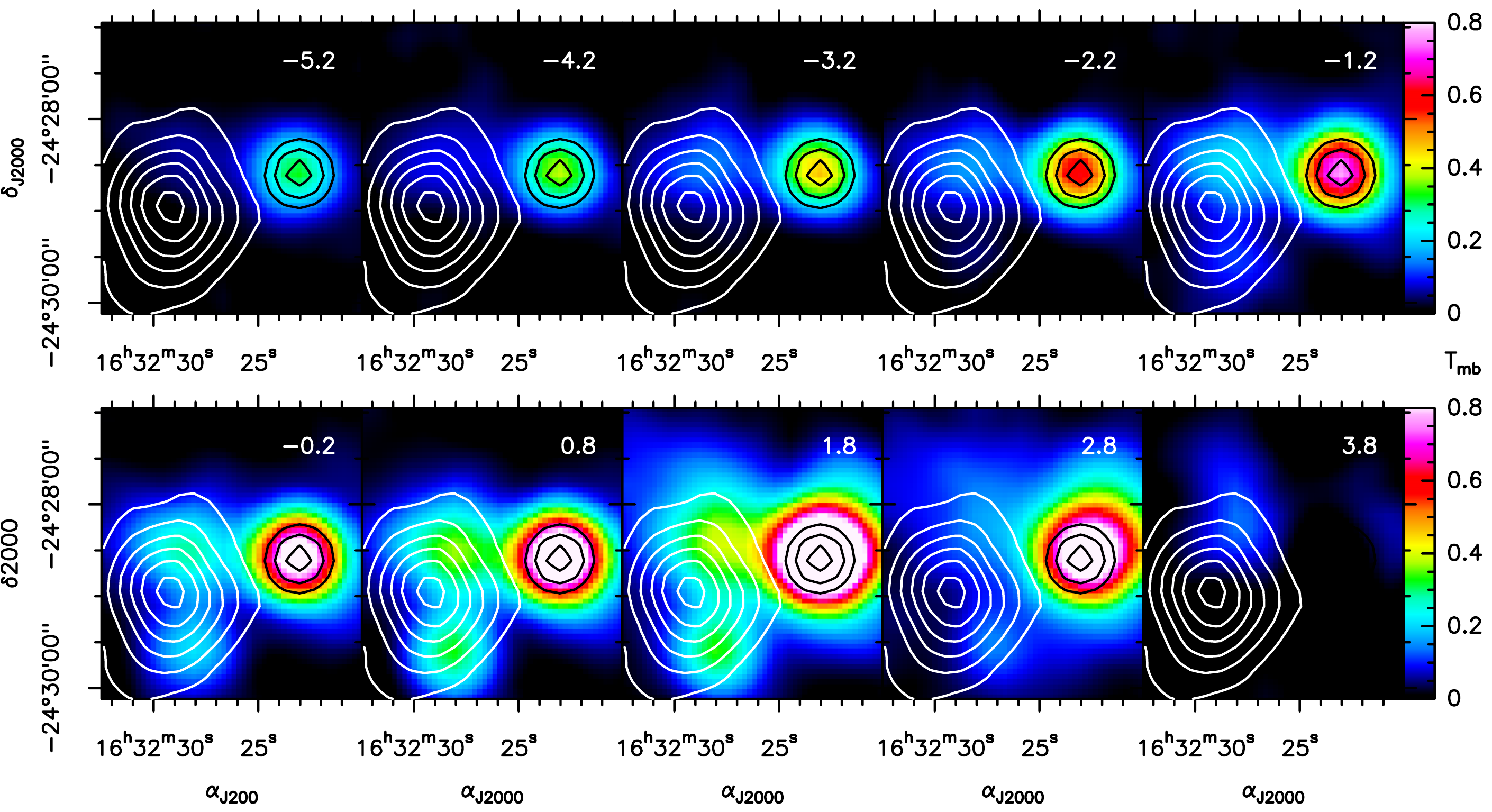}
\caption{Channel maps of the blueshifted 557~GHz water emission,
  between -5.2 and 3.8~\kms\ with overlaid white contours of the
  \nhhd\ integrated line intensity between 1.7 and 5.5~\kms, showing
  the location of the prestellar core and black contours of the
  excited SO emission showing the location of the protostar
  IRAS~16293. Contour levels are the same as in
  Fig.~\ref{fig:morphology}.}
\label{fig:channels}
\end{figure*}

At 340~GHz, a ten point mosaic covering the same \about
$65 \times 80$\arcs\ region was observed on six occasions in 2014
April with the Morita Array component of ALMA with 15 minutes of
on-source integration time per mosaic pointing. Eleven 7 m antennas
were used for the observations, with a maximum baseline of 50~m. The
0.85~mm dust continuum flux was extracted by averaging channels free
of molecular line emission. The rms is 3.3~mJy, a factor of 3 higher
than that in the 970~\mic\ continuum image, and the synthesized beam
size is $4.4 \times 2.9$\arcs, at a position angle of 94\deg. The peak
flux density of the 850~\mic\ dust continuum is 72~mJy/beam and the
integrated flux is 730~mJy. The 970~\mic\ continuum image thus has a
much higher SNR, and better preserves the low-level extended emission.
Therefore, we use the 970~\mic\ data in the subsequent analysis.

\begin{table}[!t]
\small
\begin{center}
\caption{Results of the HFS fits.}
\vskip 2mm 
\label{tab:hfs}
\begin{tabular}{lcccc}
\hline\hline
Line       & $v_0$         & $\Delta v$ & $T_{ex}$ & $\tau_0$  \\
             & (\kms)        & (\kms)       &  (K) \\ 
\hline
\nhhd\      &  3.34 & 0.38 & 6.5 & 2.7 \\
\nddd\      &  3.45 & 0.40 & 6.5 & 0.43 \\
\nndp\      &  3.68 & 0.34 & 7.9 & 3.3 \\
\hline
\end{tabular}
\end{center}
Formal $1\sigma$ fit uncertainties provided by CLASS are \about 0.002~\kms\ for the line velocity
and width, and \about 2.5\% for the line center optical depth.
\end{table}

ALMA/NAASC staff performed bandpass calibration with the quasar
J1733-1304 or J1427-4206, and the flux was calibrated with Titan. The
phase calibrator was J1626-2951. The uncertainty in the absolute flux
is ∼10\%. The calibrated visibilities were deconvolved and CLEANed
with the CASA software package (version 4.2). 

Single-dish observations of the 1~mm molecular transitions 
were carried out in 2013 May--June, using the 10.4 m Leighton
Telescope of the Caltech Submillimeter Observatory (CSO) on Mauna Kea,
Hawaii. We used the wideband 230~GHz facility SIS receiver and the
FFTS backend that covers the full 4 GHz intermediate frequency (IF)
range with a 270~kHz channel spacing (0.37~\kms\ at 220~GHz). Pointing
of the telescope was checked by performing five-point continuum scans
of planets and strong dust continuum sources. The CSO main-beam
efficiency at 230~GHz at the time of the observations was determined
from total-power observations of planets to be $\sim 65\%$. The
absolute calibration uncertainty of the individual measurements is
$\sim 15\%$. The FWHM CSO beam size is \about 35\arcs\ at 220~GHz,
comparable to the Band 1 HIFI beam size.

\section{Results}\label{sec:results}

\subsection{\emph{Herschel} Space Observatory}

The overall morphology of the region, as observed with
\emph{Herschel}, is shown in Figure~\ref{fig:morphology}. The color
image shows the integrated intensity of the 557~GHz water emission
between --15 and 25~\kms. The location of IRAS~16293 is marked by the
black contours of the integrated intensity of high-energy SO line
emission, observed simultaneously, while the prestellar core is
outlined by the white \nhhd\ contours. There is an anti-correlation
between the \nhhd\ and water emission, which surrounds and avoids the
prestellar core. This can be seen even more clearly in the velocity
channels maps (Figure~\ref{fig:channels}) at blueshifted velocities.
The strongest peak of the \hho\ emission is seen toward IRAS~16293.
However, two secondary peaks are present, to the north-west and
south-west of the prestellar core. The blueshifted emission toward the
northern peak extends to \about --8~\kms, while the emission toward
the south-western peak stops at \about --2~\kms. Water is generally
considered an excellent tracer of molecular outflows, because water
molecules are mostly frozen on dust grains in the cold gas and the
spectra are less confused by the foreground envelope absorption or
emission, compared to other tracers. We note that water vapor was
detected in the isolated prestellar core L1544 in a very long HIFI
integration \citep{caselli12}. A low gas-phase water abundance in this
source is maintained by UV photons locally produced by the interaction
of \hh\ molecules with galactic cosmic rays.

\begin{figure}[!t]
\centering
\includegraphics[width=0.9\columnwidth,angle=0]{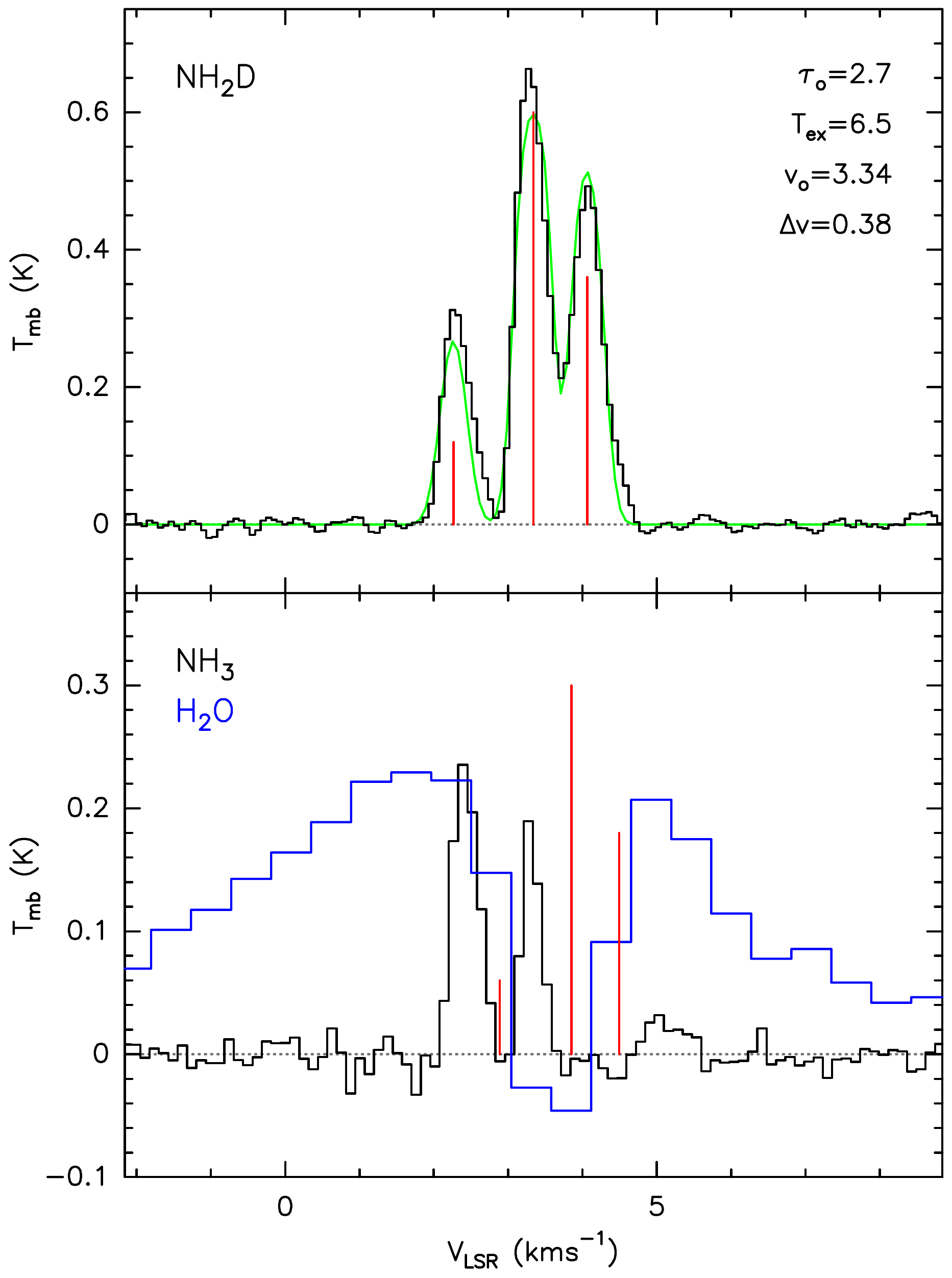}
\caption{Spectra of the \nhhd\ and \nhhh\ lines toward the prestellar
  core ($\alpha_{J2000}=16^{\rm h}32^{\rm m}29.40^{\rm s}$,
  $\delta_{J2000}=-24^{\circ} 28^{\prime} 53^{\prime\prime}$) taken
  with the HRS spectrometer. The green line in the upper panel is the
  HFS fit, which gives the line center optical depth of $2.7 \pm 0.07$
  and the excitation temperature of $6.5 \pm 0.8$~K. The \hho\
  spectrum observed with the WBS is shown in blue in the lower panel.
  Red vertical lines marked the velocities of the \nhhd\ and \nhhh\
  HFS components, with the relative line intensities corresponding to
  optically thin LTE emission (5:3:1). The \nhhh\ HFS components are
  plotted with respect to the velocity of the foreground cloud
  envelope at 3.85~\kms, as determined from the \nhhh\ absorption
  spectrum toward IRAS 16293. }
\label{fig:spectra}
\end{figure}

A high-resolution spectrum of the \nhhd\ line toward the prestellar
core is shown in Figure~\ref{fig:spectra} (upper panel). The hyperfine
structure (HFS) is well resolved spectrally, given the narrow line
width of 0.38~\kms. The HFS fit parameters (the line center velocity
and width, the excitation temperature, and the line center optical
depth) are given in Table~\ref{tab:hfs}. The weakest hyperfine
component is optically thin ($\tau \sim 0.5$). The spectrum of \nhhh\
(lower panel) looks very peculiar, as only the weakest HFS component
is clearly seen, with an intensity lower than that of the weakest
\nhhd\ hyperfine component. The \nhhh\ emission is clearly
sub-thermally excited, and the resulting spectrum is affected by the
strong temperature and density gradients along the line of sight. A
likely explanation of the peculiar line shape is that the two
strongest \nhhh\ hyperfine components are almost completely absorbed
by the foreground gas. The \nhhh\ spectrum toward IRAS~16293 shows
absorption at \about 3.85~\kms\ (red vertical lines in
Fig.~\ref{fig:spectra}, lower panel). Since the molecular emission
toward the prestellar core is blueshifted with respect to the envelope
velocity, the weakest, blueshifted \nhhh\ hyperfine component is much
less affected by the foreground absorption. An alternative explanation
that the strongest \nhhh\ emission feature corresponds to the main HFS
component at a velocity of \about 2.5~\kms, and that the weakest HFS
component is simply not detected, can be discarded as there are no
other molecular tracers emitting at 2.5~\kms\ in this source.
Due to the contamination by the
foreground absorption, the optical depth of the \nhhh\ line cannot be
directly determined from the HFS fit and the line is thus not useful
for quantitative determination of the molecular column density and the
isotopic D/H ratio.

\begin{figure*}[!t]
\centering
\includegraphics[width=0.7\textwidth,angle=0]{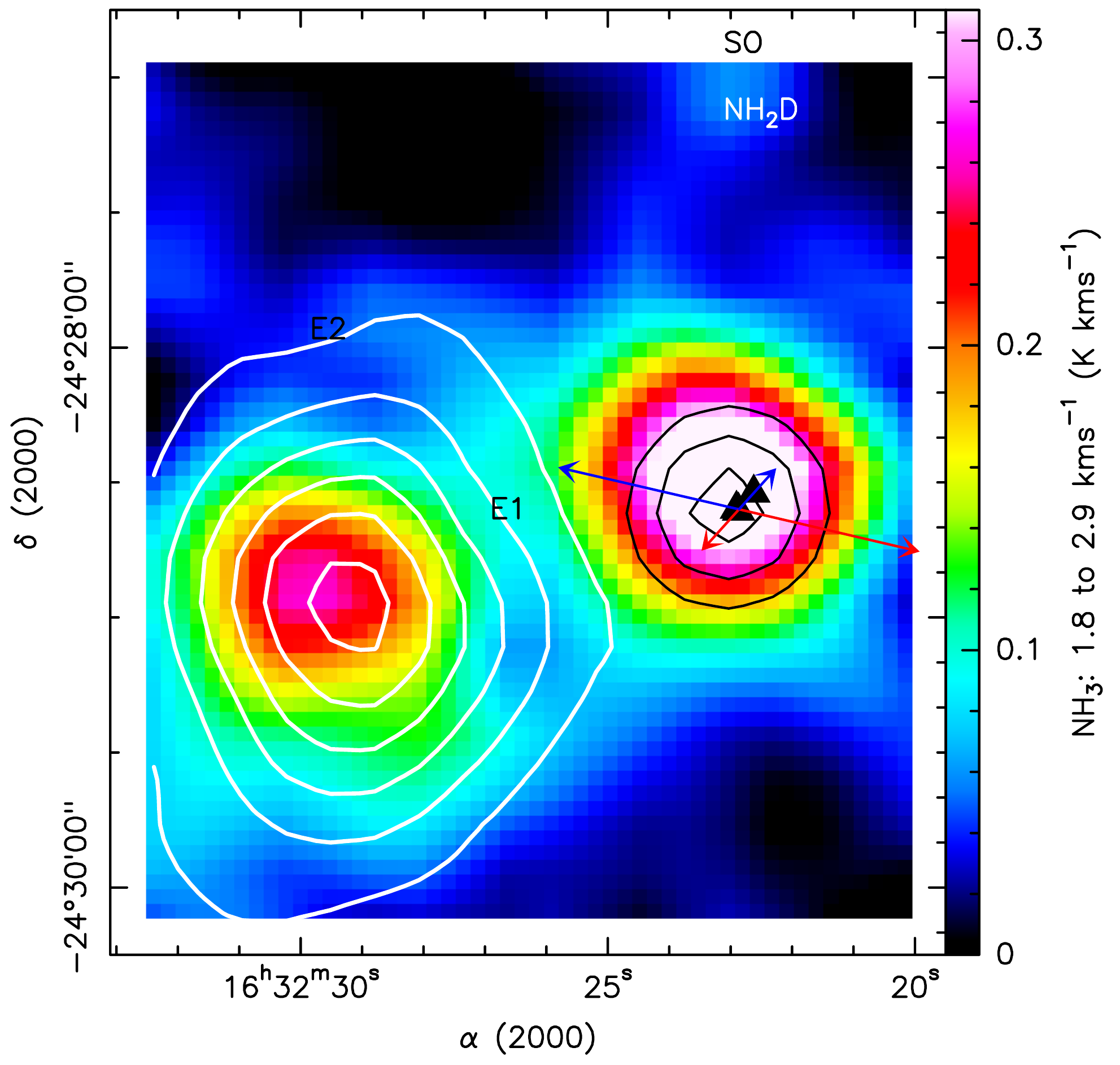}
\caption{Color image of the integrated line intensity of the weakest
  \nhhh\ HFS component, between -0.8 and 1.9~\kms, with overlaid
  contours of \nhhd\ and SO (white and black, respectively).
  Contour levels and symbols are the same as in Figure~1.}
\label{fig:ammonia}
\end{figure*}

Figure~\ref{fig:ammonia} shows the integrated line intensity of the
weakest \nhhh\ HFS component, with overlaid contours of the \nhhd\ and
SO emission. In the vicinity of the prestellar core, \nhhh\ peaks
\about 10\arcs\ to the east of \nhhd. The offset may be
indicative of an abundance gradient between the two ammonia
isotopologues. However, given the high opacity of the \nhhh\ line, it
may simply be due to radiative transfer effects.

\begin{figure*}[!t]
\centering
\includegraphics[width=0.9\textwidth,angle=0]{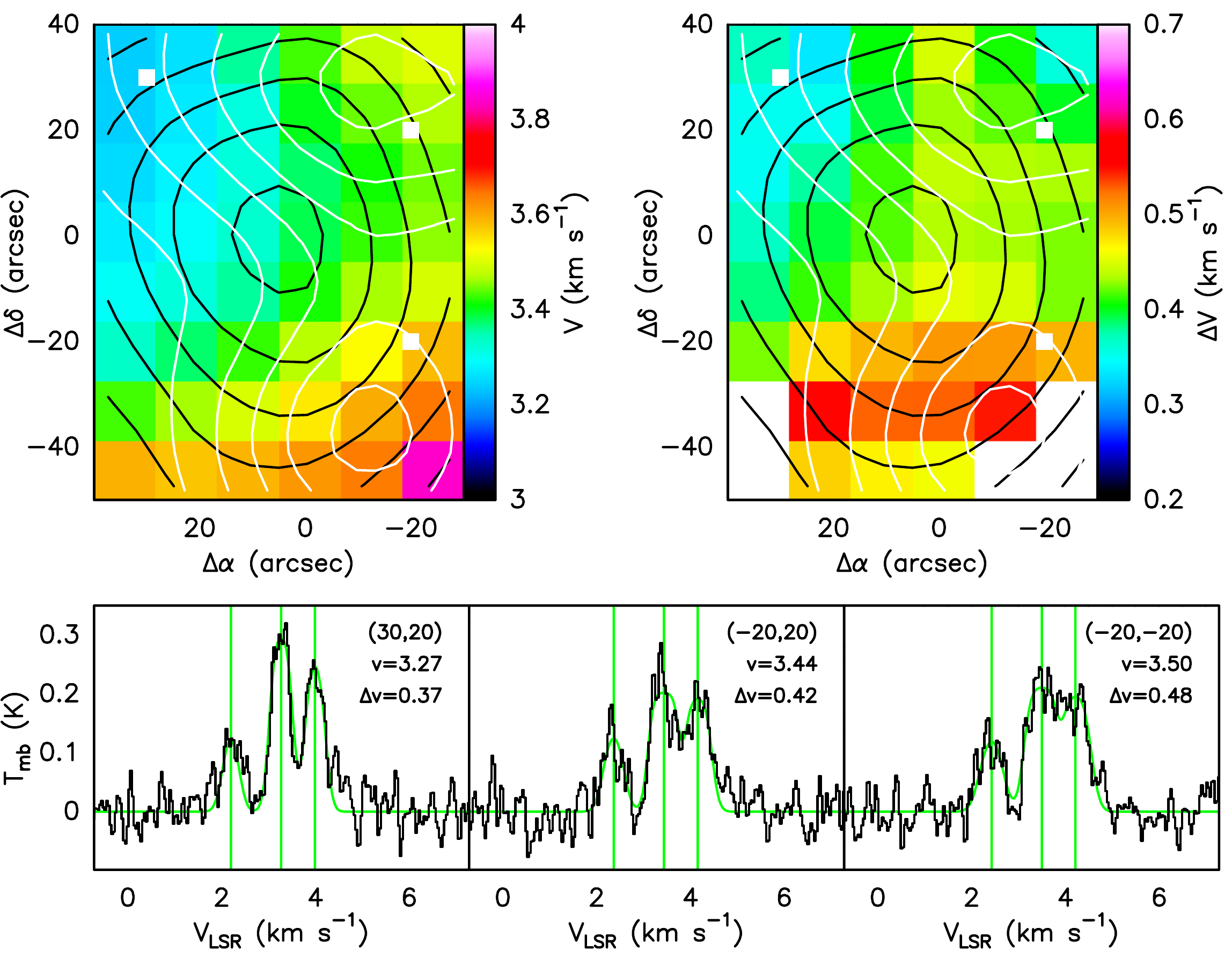}
\caption{(Top) Maps of the \nhhd\ line velocity and width across the
  prestellar core (left and right, respectively). Black contours show
  the \nhhd\ integrated line intensity and white contours blue-shifted
  \hho\ emission, between --0.2 and 3.8~\kms. Contour levels are
  from 20\% to 95\% of the peak (0.65 and 1.11~K\,\kms\ for \nhhd\ and
  \hho, respectively), with a step of 15\%. The (0,0) position
  corresponds to $\alpha_{J2000}=16^{\rm h}32^{\rm m}28.84^{\rm s}$,
  $\delta_{J2000}=-24^{\circ} 28^{\prime} 57^{\prime\prime}$. (Bottom)
  \nhhd\ spectra at three positions across the prestellar core, marked
  by white squares in the top row, demonstrating the change in the
  line velocity and width from the north-east to the south-west. The
  spectra have been convolved with a 25\arcs\ Gaussian to improve the
  SNR. }
\label{fig:nh2dvelo}
\end{figure*}

An interaction with the outflow should have an effect on the velocity
field of the prestellar core. Figure~\ref{fig:nh2dvelo} shows the
\nhhd\ line center velocity and width as a function of position (top
left and right panels, respectively). There is a clear trend of the
line velocity and width increasing from the north-east to the
west/south-west. The \nhhd\ spectra at three selected positions, shown
in the bottom row, display this transition from a quiescent gas with a
narrow line width in the north-east, to a much more perturbed gas in
the western/south-western part of the core.

\subsection{ALMA Compact Array}\label{sec:aca}

\begin{figure*}[!t]
\centering
\includegraphics[width=0.9\textwidth,angle=0]{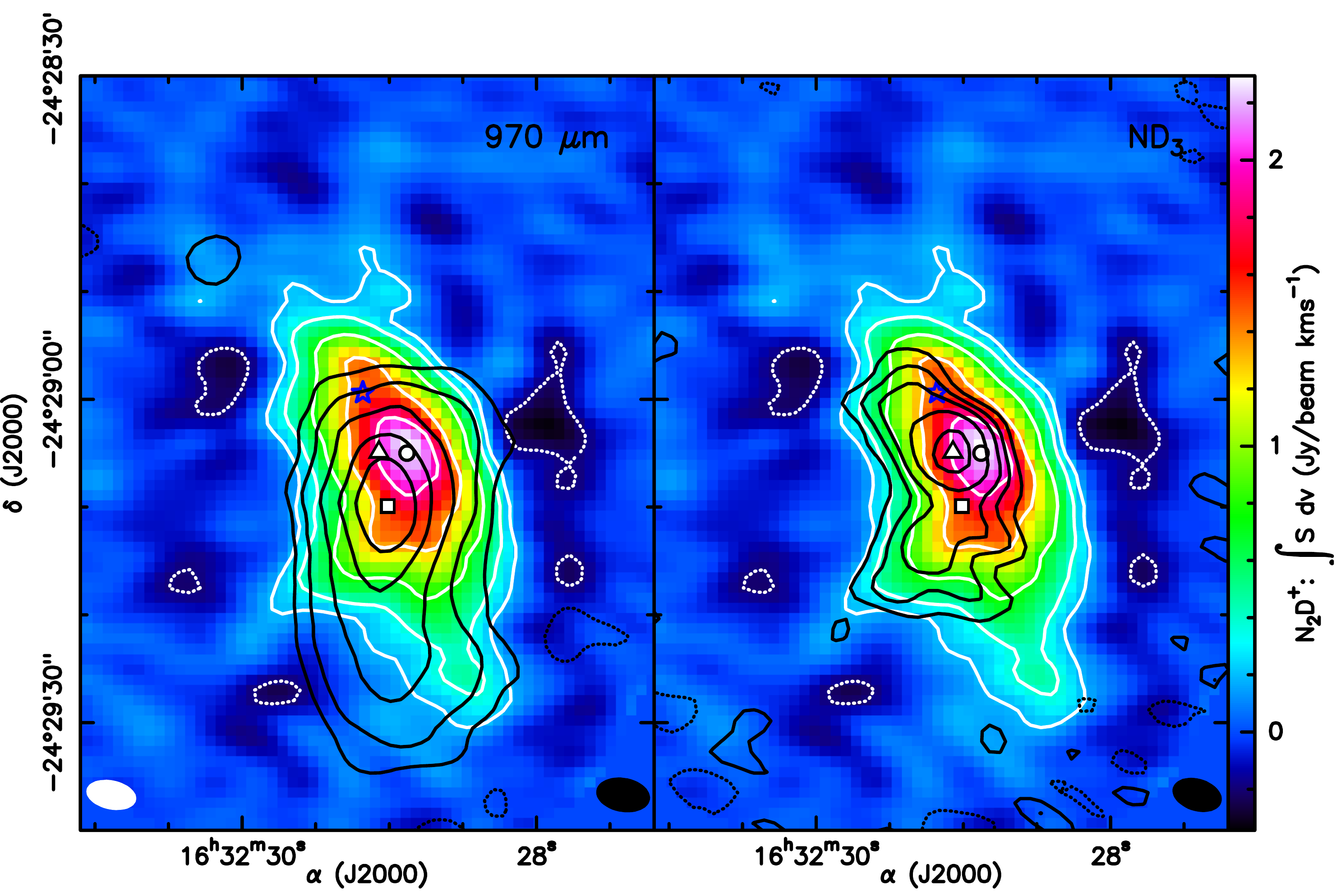}
\caption{Integrated line intensity of the \nndp\ emission (color
  image and white contours), with overlaid black contours of the
  970~\mic\ continuum emission (left panel) and integrated line
  intensity of the \nddd\ emission (right panel). \nndp\ emission has
  been integrated over velocities 3.04--4.17 and 5.24--5.77~\kms,
  including 2 groups of the strongest hyperfine components, while the
  \nddd\ emission has   been integrated over velocities 3.08--3.79 and
  4.32--4.97~\kms,  including the 2 strongest hyperfine components
  (including the weaker hyperfine components does not improve the SNR
  in the resulting images).  Contour
  levels for the \nndp\ and the continuum emission are: --5, 5, 10, 20,
  35, and 50 times the rms  ($3.84\times 10^{-2}$~Jy/beam\,\kms\ and
  $1.11\times 10^{-3}$~Jy/beam, respectively). For \nddd, contour
  levels are: --3, 3, 5, 7, 10, and 14 times the rms ($3.67\times
  10^{-2}$~Jy/beam\,\kms). The synthesized beams are: $4.71 \times
  2.69$\arcs\ at 78\deg\ for \nndp\ (shown in the lower-left corner of the
  left panel), $4.70\times 3.00$\arcs\ at 73\deg\
  for \nddd\ (lower-right corner of the right panel), and
  $5.04\times 3.16$\arcs\ at 80\deg\ for 970~\mic\ continuum
  (lower-right corner of the left panel). Symbols mark
  the location of the emission peaks: \nndp\ (circle; 16\hr 32\mt 28\secper 88,
  --24\deg 29\arcm 05\arcsper 0), \nddd\ (triangle; 16\hr 32\mt
  29\secper 07, -24\deg 29\arcm 04\arcsper 9), continuum
  (square; 16\hr 32\mt 29\secper 01, -24\deg 29\arcm 09\arcsper 9),
  and \nhhd\ (star; 16\hr 32\mt 29\secper 2 -24\deg 28\arcm
  59.4\arcs). 
}
\label{fig:alma1}
\end{figure*}

To further investigate the morphology of the prestellar core, we have
imaged the 970~\mic\ dust continuum emission, together with a number
of molecular tracers, including deuterated species, using the ALMA
Compact Array. The \nndp\ emission (Figure~\ref{fig:alma1}, color
image and white contours) extends south/south-west from the peak of
the \nhhd\ emission, as observed with HIFI (blue star), and peaks
\about 5\arcs\ to the north with respect to the dust continuum (black
contours, left panel). While the northern tip of the \nndp\ emission
coincides with the \nhhd\ peak, the very large 44\arcs\ HIFI beam
prevents a detailed comparison of the morphology on the angular scales
probed by the ACA. We note, however, that the dust emission terminates
rather abruptly in this part of the core, with the molecular emission
extending further to the north. The \nddd\ emission (black contours,
right panel), resembles more closely \nndp\ than the dust continuum
emission, although the \nddd\ emission peak is displaced by \about
2.5\arcs\ east with respect to \nndp, (possibly due to optical depth
effects, as the \nndp\ line has a significant opacity,
Table~\ref{tab:hfs}).

\begin{figure*}[!t]
\centering
\includegraphics[width=0.7\textwidth,angle=0]{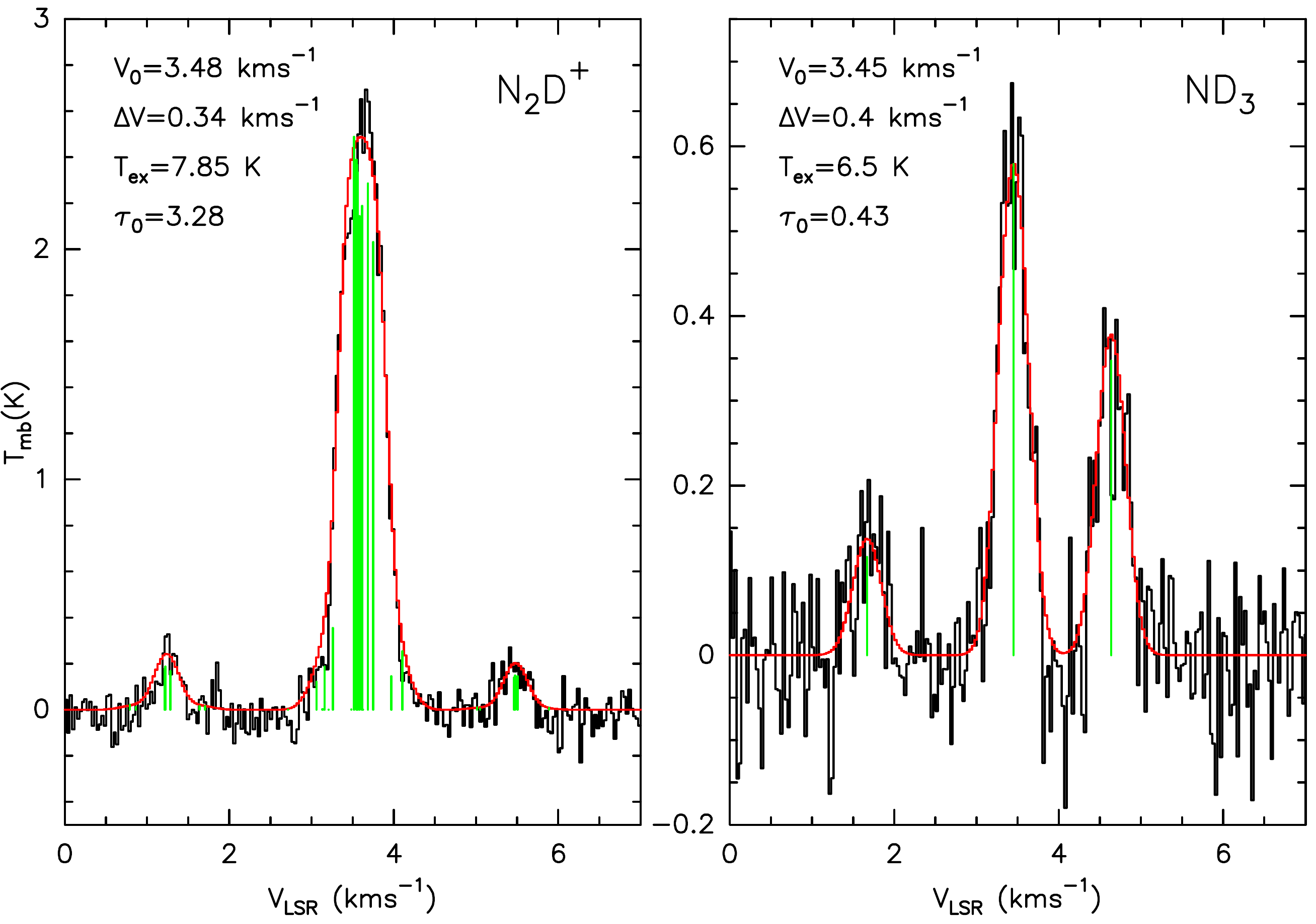}
\caption{\nndp\ and \nddd\ spectra (left and right panels,
  respectively) toward the respective emission peaks. HFS fits are
  shown in red with the best fit parameters shown in the upper-left
  corners. For \nndp\ the LSR velocity scale is plotted with respect
  to the CDMS frequency of 308422.27~GHz.}
\label{fig:almafit}
\end{figure*}
 
Figure~\ref{fig:almafit} shows \nndp\ and \nddd\ spectra toward the
respective emission peaks, with single velocity component HFS fits
overlaid in red. Vertical green lines mark the velocities and relative
intensities of the HFS components. The HFS fit parameters are given in
Table~\ref{tab:hfs}. The \nndp\ emission is quite optically thick and
does not trace directly the molecular column density. The derived
excitation temperature, which can be taken as a lower limit of the gas
kinetic temperature (the emission does not have to be thermalized), is
significantly lower that the isothermal dust temperature of 16~K
derived from a SED fit by \cite{stark04}. However, these authors point
out that a cold, high-density region with $T_d \ll 16$~K can easily be
hidden at the center of the core. In fact, in a recent study,
\cite{bacmann16} derive a temperature profile increasing from 11~K at
the core center to 16~K in the outer envelope. Even lower dust
temperatures, in the range 8--16~K, are derived by \cite{pagani16}.

The \nddd\ emission is optically thin. A HFS fit gives a slightly
larger line width compared to \nndp\ (Table~\ref{tab:hfs}), which
however is consistent with that of the \nhhd\ emission in the ten
times larger HIFI beam. We thus see no evidence in our deuterated
ammonia data for the turbulent velocity to vary with radius, as seen
in some sources. The complex HFS of \nndp\ and a limited SNR of the
\nddd\ spectrum prevent us from drawing any conclusions about the
presence of possible infall motions in the core. The \about 0.4~\kms\
FWHM line width roughly corresponds to the \hh\ thermal line width at
7~K and is 3--4 times larger than the expected thermal line width of
\nddd, \nhhd, or \nndp. This shows that the line broadening is mainly
non-thermal and that sonic or somewhat sub-sonic turbulent motions are
dominant even in the northern, quiescent part of the core. This is
different from the typical prestellar cores in Taurus, where line
widths are essentially thermal and is perhaps related to the
interaction with the outflow.

\begin{figure*}[!t]
\centering
\includegraphics[width=0.9\textwidth,angle=0]{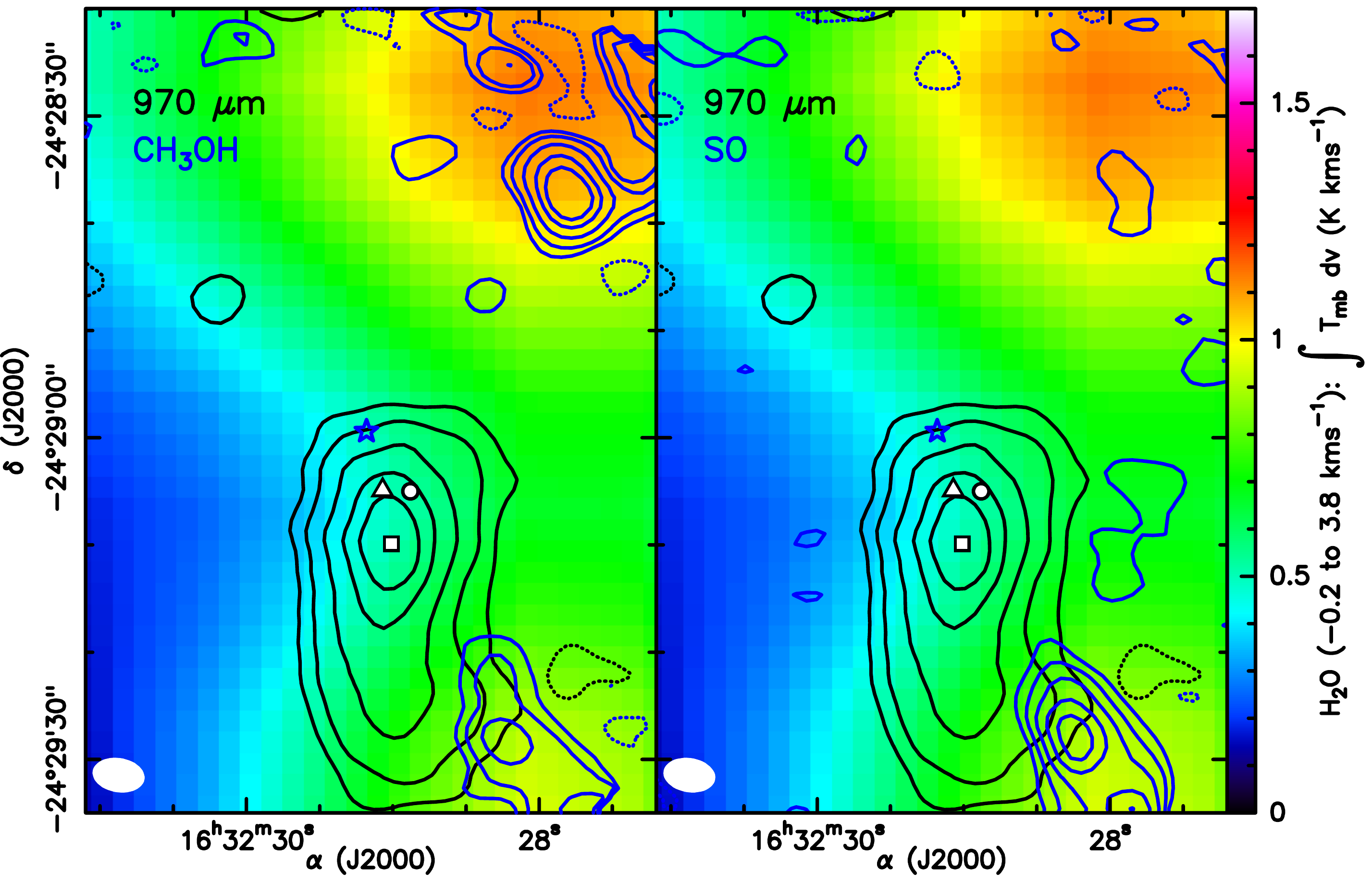}
\caption{Integrated line intensity of the blueshifted \hho\ emission
  between --0.2 and 3.8 \kms\ (color image), with overlaid black
  contours of the 970~\mic\ continuum emission and blue contours of
  the integrated line intensity of the CH$_3$OH and SO emission
  (left and right panels, respectively). \nndp\ emission has been
  integrated over velocities --2.56 to 5.49~\kms, while the SO
  emission has been integrated over velocities 0.90 to 4.21~\kms.
  Contour levels for the CH$_3$OH and the continuum emission are: --5,
  5, 10, 20, 35, and 50 times the rms
  ($6.8\times 10^{-2}$~Jy/beam\,\kms\ and
  $1.11\times 10^{-3}$~Jy/beam, respectively). For SO, contour levels
  are: --3, 3, 6, 10, 15, and 20 times the rms
  ($6.7\times 10^{-2}$~Jy/beam\,\kms). The synthesized beam for the
  CH$_3$OH and SO images is $4.88 \times 3.19$\arcs\ at 81\deg\ for
  \nddd\ (lower-right corner of the right panel), and
  $5.04\times 3.16$\arcs\ at 80\deg\ for 970~\mic\ continuum (shown in
  the lower-right corner). Symbols are the same as in Fig.
  ~\ref{fig:alma1}. }
\label{fig:alma2}
\end{figure*}

Figure~\ref{fig:alma2} shows the distribution of the CH$_3$OH and SO
emission in the vicinity of the prestellar core (blue contours, left
and right panels, respectively). The background image is the
blueshifted \hho\ emission observed with HIFI, while the black
contours show the 970~\mic\ dust continuum emission. A significant
fraction of the CH$_3$OH and SO emission (\about 80--85\%) is resolved
out in the ACA images. We see however, strong methanol emission
associated with the two blue-shifted water peaks to the north-west and
south-west of the prestellar core, while the SO emission is only
associated with the weaker south-western blueshifted water peak.
\cite{pagani16} report the presence of a new outflow, which could be
responsible for the emission peak south-west of the prestellar core.
No SO or methanol emission is detected toward the prestellar core.

\subsection{Caltech Submillimeter Observatory}

\begin{figure*}[!t]
\centering
\includegraphics[width=0.9\textwidth,angle=0]{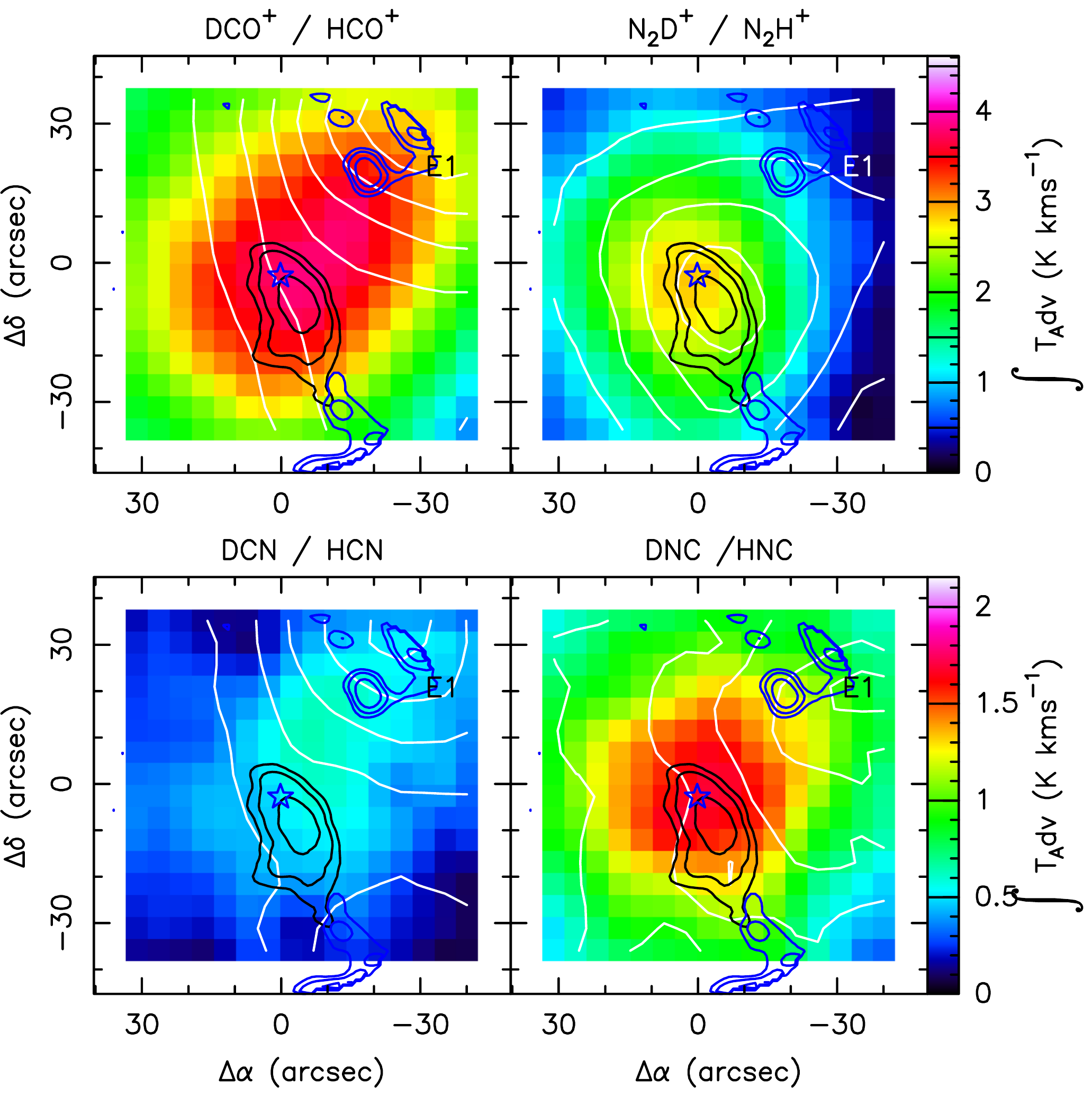}
\caption{CSO maps of deuterated species in L1689N (color images), with
  overlaid white contours of their hydrogenated counterpars. \nndp\
  and DNC are the best tracers of the prestellar core, with \nndp\
  being more centrally peaked. Both \dcop\ and DCN emission show
  extensions toward the E1 source in the north-west. Black contours
  show the distribution of \nhhd, as observed with the ACA, while the
  blue contours show the distribution of CH$_3$OH. Contour levels are:
  5 to 17.5 with a step of 2.5~K\,\kms\ for \hcop; 2.5 to 7 with a
  step of 1.5~K\,\kms\ for \nnhp; 2 to 8 with a step of 1.5~K\,\kms\
  for HCN; 1 to 3 with a step of 0.5~K\,\kms\ for \hcop. ACA contour
  levels are 10, 20, 40 times the rms. Star marks the location of the
  \nhhd\ emission peak observed with HIFI.}
\label{fig:csomaps}
\end{figure*}

To study further the large-scale morphology and kinematics of the
molecular gas in the vicinity of the prestellar core, we have mapped
the 1~mm lines of \dcop, DCN, DNC, and \nndp, as well as their
hydrogenated counterparts \hcop, HCN, HNC, and \nnhp, using the CSO
(Table~1).\footnote{The \nnhp\ data have been previously used in the
  models of \cite{daniel16a} to derive the nitrogen isotopic ratio in
  L1689N. } The resulting maps of deuterated species are shown as
color images in Figure~\ref{fig:csomaps}. Although all deuterated
species peak in the general vicinity of the prestellar core,
significant differences among their morphologies can be seen. The
prestellar core, as traced by the \nhhd\ emission, is best seen in the
DNC and \nndp\ images. The \dcop\ emission, being the brightest and
thus most suitable for mapping of extended areas, is not confined to
the immediate vicinity of the prestellar core and shows a clear
extension toward the E1 shocked region, while DCN peaks in-between the
prestellar core and E1. This is not surprising, as this molecule has
been suggested to be produced by ``warm chemistry'' driven by
CH$_2$D$^+$, in the gas with temperatures of order 50 K, where the
H$_2$D$^+$ driven deuteration reactions are already suppressed
\citep{roueff07, parise09}. We note that, similarly to \nhhd, the
single dish \nndp\ and DNC emission also peaks to the north of the ACA
\nndp\ and \nddd\ sources. Among the hydrogenated molecular tracers,
only \nnhp\ emission reveals the prestellar core, peaking close to the
location of the ACA \nndp\ and \nddd\ sources. All the other tracers
largely follow the distribution of the blueshifted water emission,
peaking toward the E1 source, with a secondary peak to the south-west
(labeled HE2 by \citealt{castets01}).

\section{Discussion}\label{sec:discussion}

\subsection{Properties of the Dust Continuum Source}


The 970~\mic\ continuum source detected with the ACA is elongated
approximately in the north-south direction. A two-dimensional Gaussian
fit gives a FWHM source size of $13.2 \times 8.6$\arcs\ at a position
ange of 2\deg. After deconvolving the beam, the intrinsic source size
is $12.9 \times 7.0$\arcs, or $1540 \times 840$~au, assuming a
distance of 120~pc \citep{loinard08}. The column density profile of
the core is well described by a two-dimensional Gaussian.
However, the spatial dynamical range in the current image is rather
low, only \about 3.5. Higher angular resolution observations with the
main ALMA array will thus be required to study this important aspect.

The peak 970~\mic\ continuum flux is 76.0~mJy/beam and the integrated
flux is 0.84~Jy. To estimate the \hh\ column density and the mass of
the compact dust source, we scale the grain opacity coefficient
$\kappa_{\rm 1.3 mm} = 0.005$~cm$^2$ g$^{-1}$ \citep{motte98},
appropriate for prestellar dense, clumps with a $\nu^2$ frequency
dependence. This leads to $\kappa_{\rm 970 \mu m} = 0.009$~cm$^2$
g$^{-1}$.\footnote{\cite{stark04} used a grain opacity coefficient
  $\kappa_{\rm 1.2 THz}=0.1$\pscm g$^{-1}$. When extrapolated with a
  $\nu^2$ frequency dependence, this gives
  $\kappa_{\rm 870 \mu m} = 0.008$~cm$^2$ g$^{-1}$, about 25\% lower
  than the value used here.} For a 16~K dust temperature
\citep{stark04}, we derive an \hh\ column density of
$1.8 \times 10^{23}$~\pscm. However, if the dust temperature is 11~K,
as suggested by \cite{bacmann16}, the corresponding column density
increases to $3.3 \times 10^{23}$~\pscm. Even in the latter case, the
dust emission is optically thin (a 970~\mic\ optical depth of \about
0.01). Assuming a $\nu^2$ frequency dependence, the dust opacity at
557~GHz is also low. Therefore, the observed anti-correlation between
water and \nhhd\ emission, as seen by HIFI, cannot be explained by
foreground dust absorption.

Assuming a line-of-sight source size of 9.5\arcs\ or 1140~au, equal to
the geometric mean of the FWHM sizes in the plane of the sky, we
derive an \hh\ volume density of $(1.1-1.9)\times 10^7$~\pccm,
depending on the dust temperature. This is an order of magnitude
higher than the values derived in earlier single-dish studies
\citep{stark04, vastel12}, but consistent with the best-model value of
$1.4 \times 10^7$~\pscm\ in the recent study of \cite{bacmann16}. The
high central density in the model of \cite{bacmann16} is in fact
directly related to the presence of a temperature gradient. The
best-fit power-law density exponent in these latest models is 1.7, as
compared to the low value of \about 1 in the isothermal model of
\cite{stark04}. The prestellar core can also be identifies with source
84 (SMM 19) of \cite{pattle15}, who derive a temperature of 11.8~K and
a density of $3.3\times10^7$~\pccm. The new single-dish and
interferometric data presented here provide additional constrains for
a detailed radiative transfer modeling, which will be subject of a
separate study.

The total mass of the compact continuum source seen in the ACA image
is 0.22~M\solar, assuming a dust temperature of 16~K (0.42~M\solar\ for
a dust temperature of 11~K), a small fraction of the total mass of the
prestellar core 
2.45~M\solar, \citealt{stark04} (assuming a 16~K dust temperature and
scaling to the distance of 120~pc used here). 
\cite{pattle15} derive a mass of 0.30~M\solar\ (after scaling to the
distance of 120~pc). This source thus has a mass characteristic of the
Very Low Luminosity Objects (VeLLOs) detected by \emph{Spitzer} (see,
e.g., \citealt{dunham14} and references therein), which have
bolometric luminosities below 0.1~L\solar\ \citep{difrancesco07}.
\cite{stark04} derive a total bolometric luminosity of
1.3~L\solar\ toward the prestellar core in L1689N, after scaling to the distance
of 120~pc. However, it is not clear what fraction of this luminosity is
associated with the compact continuum source detected in the ACA data,
as \emph{Spitzer} IRAC images do not reveal any 4.5~\mic\ continuum
sources at this location \citep{pagani16}.

\subsection{Interaction with the Outflow}

The environment of L1689N shows a complex morphology. Optically
thick molecular lines show a pronounced self-reversal at the systemic
velocity, \about 3.8~\kms. The parent molecular cloud has a mean density of
$(2-3) \times 10^4$~\pccm\ and a kinetic temperature of 12~K, based
on ammonia observations \citep{menten87}. Multiple large-scale
molecular outflows are present, driven by IRAS~16293 \citep{mizuno90,
  lis02a, stark04}. The SMA observations of \cite{yeh08} imply a
primarily east-west outflow, with blueshifted emission to the west.
These authors suggest that this small-scale outflow may be the inner
part of the large-scale east-west outflow. More recent SMA data
\citep{girart14} suggest that the CO emission at moderately high
velocities arises from two bipolar outflows, which appear
perpendicular to each other and both originate from IRAS~16293A. The
more compact outflow is impacting the circumstellar gas around
component B, and is possibly being redirected. The directions of the two
compact outflows, as indicated by the SMA observations of
\cite{girart14}, are marked in Figure~\ref{fig:morphology}.

The outflow originates in the two sources at the center of
IRAS~16293A, A-E and A-W. In fact, VLA observations reveal that the
continuum source A-E orbits A-W, which had a recent continuum ejection
event \citep{loinard07}. That event defined the flow originating in
A-W, with the redshifted emission to the south-west. VLBI water maser
observations agree with this picture, showing a very well-defined
bow-shock among the redshifted masers to the south-west (A. Wootten,
private comm.) Source A-E, in orbit around A-W, also shows maser
outflow, approximately parallel to that in A-W and in the same sense,
i.e. redshifted emission to the south-west. On larger scales the
overall complexity of the L1689N region makes the interpretation of
the outflow difficult. However, one clear conclusion is that there is
a lot of moderate-velocity blueshifted gas in the immediate vicinity
of the prestellar core and the change in the observed \nhhd\ line
velocity and width is a direct evidence that this gas is
interacting with the prestellar core.

Interestingly, the polarity of both compact CO outflows seen in the
SMA data \citep{girart14} seems opposite to that of the large-scale
outflows seen in the earlier lower-resolution single-dish data
\citep{mizuno90, lis02a, stark04} (compare, e.g., Figure 1 of
\citealt{girart14} and Figure 17 of \citealt{stark04}). The northern
peak of the blueshifted water emission appears to be a part of the
main east-west outflow detected in the SMA CO data. VLBA observations
of the water maser proper motions \citep{wootten99} also show that the
IRAS~16293A outflow proceeds red NE, blue SW, from within a few AU of
the sources. 

The key observational aspects of the new data presented here that have
to be satisfactorily explained by a source model are: (a) the change
in the \nhhd\ line center velocity and width across the core
(Figure~\ref{fig:nh2dvelo}); (b) the apparent offset between the peak
of the 970~\mic\ dust continuum emission and the emission of
deuterated molecular tracers in the ACA data (Figure~\ref{fig:alma1}).

One possible explanation is that the prestellar core in L1689N is
simply a pre-existing density enhancement, that had formed
independently of IRAS~16293. This region is characterized by some of
the highest deuteration levels known in galactic sources. For example,
it is one of a handful of sources where triply deuterated ammonia has
been detected \citep{roueff05}. Such cores are generally rare, but
Barnard~1 offers another example of a similar region. Here, a dense
prestellar core, B1b (also referred to as SMM1), with very high
deuteration levels, which has been suggested to host the first
hydrostatic core \citep{gerin15}, is also located in close proximity
to more evolved protostellar sources, B1a (SMM6) and B1c (SMM2),
driving pronounced outflows \citep{hatchell07, hiramatsu10}. High
deuteration levels require a low temperature, high density and
consequently high CO depletion. Such conditions are found in isolated
prestellar cores, e.g. L1544, which are characterized by a centrally
peaked density distribution that can be traced well by dust continuum
emission. However, our ACA data reveal no bright compact dust
continuum sources in the northern part of the core, where the
strongest emission of deuterated molecular tracers is seen in
single-dish maps.

\cite{lis02a} suggested that this region may be a part of the ambient
cloud that is pushed and compressed by the outflow. The
shock-compressed, dense gas then cools efficiently to low
temperatures, and the gas-phase chemistry quickly drives up abundances
of deuterated molecular species. This scenario is in fact consistent
with the observations presented here. The blue lobe of the IRAS~16293
east-west outflow impacts the prestellar core from the back. The shock
could have already propagated through the north-eastern part of the
core, where we now see narrow, undisturbed line profiles in the cold,
compressed post-shock gas, blueshifted with respect to the systemic
velocity of the ambient cloud. An order of magnitude density
enhancement, as compared to the ambient cloud, expected in the
post-shock gas, produces conditions propitious for the high
deuteration levels observed. The west/south-west part of the core is,
however, still interacting with the outflow, hence the disturbed
appearance of the molecular line profiles, with broader line widths.
The shock to the north/north-west of the core (traced by, e.g., the
SiO or CH$_3$OH emission) may be associated with a higher-velocity
component of the flow, possibly a jet or jet remnant, as the highest
blueshifted water and CO emission occurs there. There may also be a
pre-existing density enhancement at this location, which happens to
lie in the path of the higher-velocity component of the flow, which
was further compressed and heated up by the shock. This would explain
the strong emission of high-density molecular tracers, such as \hcop,
HCN, HNC, and SiO in the single dish images (Figure~\ref{fig:csomaps};
\citep{hirano01}). This region also stands out in the ACA CH$_3$OH
image and thus shows the classic outflow-dense gas interaction
signatures.

The apparent offset between the dust continuum peak, as observed with
the ACA, and the emission of deuterated molecular tracers is another
intriguing aspect. Such offsets have been observed in other prestellar
cores (e.g., L183; \citealt{tine00}), or in outflows (e.g., L1157;
\citealt{fontani14}). The dust continuum emission is typically taken
as a measure of the \hh\ column density, indicating the location of
the densest central parts of prestellar cores. However, possible
variations in the dust temperature may strongly affect this
interpretation. In fact, multi-wavelength \emph{Spitzer} and
\emph{Herschel} continuum data \citep{pagani16} indicate a NW-SE
temperature gradient across the prestellar core, with colder dust
located in the north. While the average dust temperature in a
single-dish beam is \about 16~K \citep{stark04}, this may be dominated
by the outer surface layers and the dust temperature in the central,
most shielded regions may be significantly lower (see
\citealt{bacmann16}). The ACA 850/970~\mic\ flux ratio map shows no
evidence for a temperature gradient across the compact source.
However, the SNR in our 850~\mic\ continuum image is too low to draw
definitive conclusions.

\begin{figure*}[!t]
\centering
\includegraphics[width=0.9\textwidth,angle=0]{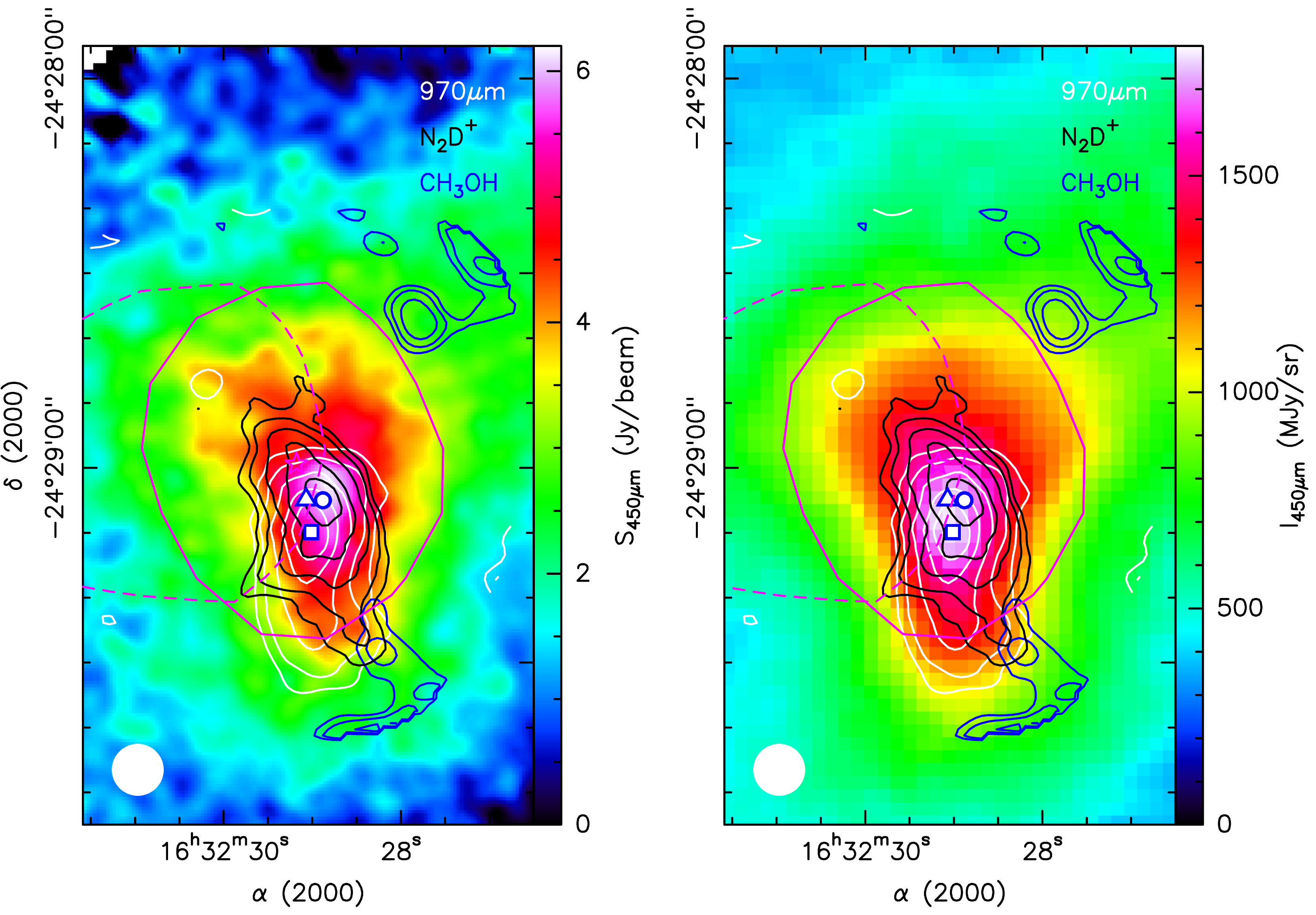}
\caption{Color images of the single-dish 450~\mic\ dust continuum
  emission toward the prestellar core, observed with SCUBA
  (\citealt{stark04}; left panel) and SCUBA-2 (\citealt{pattle15};
  right panel), with overlaid contours of the 970~\mic\ continuum,
  \nndp, and CH$_3$OH emission observed with the ACA (white, black,
  and blue contours, respectively). Contour levels are the same as in
  Fig.~\ref{fig:alma1} and \ref{fig:alma2}. The solid magenta contour
  shows the location of the \nhhd\ emission, as observed by HIFI,
  while the dashed magenta contour outlines the location of the \hhdp\
  emission \citep{pagani16} (60\% contour levels).}
\label{fig:sandell}
\end{figure*}

Another important aspect is that some of the extended continuum
emission may simply be filtered out in the ACA data.
Figure~\ref{fig:sandell} shows 450~\mic\ SCUBA and SCUBA-2 dust
continuum emission images toward the prestellar core
(\citealt{stark04, pattle15}, left and right panels, respectively),
with overlaid contours of the 970~\mic\ continuum and \nndp\ emission
observed with the ACA (green and black contours, respectively). The
single-dish dust continuum emission in the SCUBA image correlates well
with the \nndp\ emission and is clearly shifted to the north with
respect to the 970~\mic\ ACA continuum peak. The lower angular
resolution 870~\mic\ SCUBA image shows a similar offset. However, the
450~\mic\ SCUBA-2 image peaks closer to the ACA continuum peak.
The small differences between the two images may be caused by limited
SNR or pointing offsets. The integrated 345~GHz ACA flux is only
0.72~Jy, as compared to the peak 870~\mic\ broadband flux of 1.4~Jy in
the 14\arcs\ JCMT beam \citep{stark04}. The minimum baseline of the
ACA is 8.7~m, for a transiting source. The effective baseline depends
on the hour angle, but for a transiting source, the ALMA Technical
Handbook provides a value of 13.8\arcs\ for the largest angular scale
recovered by the interferometer. This strongly suggests that the ACA
continuum image may not represent the true \hh\ column density
distribution in the prestellar core. It appears instead that the
northern, quiescent part of the core is characterized by spatially
extended continuum emission, which is mostly resolved out by the
interferometer, while the southern part is more compact and centrally
peaked. This is consistent with the interpretation outlined above, in
which the northern part of the core is compressed, high-density, cold
post-shock gas, which is relatively uniformly distributed and not yet
fragmenting to form new stars. This shows that good care should be
taken when interpreting ALMA observations of extended sources, even
with the ACA, when the total power data are not included.

Interestingly, the \hhdp\ emission shows a very distinct morphology
compared to any of the tracers studied here, peaking largely to the
east/north-east (see Figure~1 of \citealt{pagani16} and the dashed
magenta contour in Figure~\ref{fig:sandell}). \cite{pagani16} argue
that this is in fact the location of the coldest, highest column
density gas in the region. We note that a weak 970~\mic\ continuum
peak, with a flux density of 6.9~mJy/beam (a 6$\sigma$ detection) is
present within the \hhdp\ emission region, which can be identified
with source 86 (SMA~22) of \cite{pattle15}. An extension in this
direction is also seen in the 870~\mic\ SCUBA image. More sensitive
observations with the main ALMA array are required to determine the
exact nature of this source.

Another important factor is that broadband, single-dish continuum
images may be contaminated by molecular line emission. As shown in
Figure~\ref{fig:csomaps}, only the classical
tracers of dense cold gas, namely deuterated species and \nnhp, peak
toward the prestellar core. Other molecular tracers largely avoid the
core and peak instead toward the E1 shock. However, a weak
\hthcop~4--3 emission peak associated with the northern part of the
core was detected by \cite{gerin06}. The blueshifted CO emission to
the north of the prestellar core may also influence the broadband
continuum images, given the relatively broad line widths. Molecular
line contamination may thus effectively shift the broadband continuum
peak to the north, away from the compact continuum source seen in the
ACA image, explaining the observed morphology. However, one clear
conclusion is that there are no strong compact continuum sources
embedded in the northern, quiescent, part of the prestellar core.

We note that \nddd\ emission was detected with a high SNR toward the
prestellar core in the single-dish CSO data \citep{roueff05} with an
integrated line intensity in the 25\arcs\ CSO beam of 0.29~K\,\kms\
(main beam brightness temperature), or \about 9.4~Jy\,\kms. The
integrated \nddd\ flux in the ACA image is 6.8~Jy\,\kms, indicating that
a significant fraction of the single dish flux is in fact recovered by
the interferometer.

\section{Summary}\label{sec:summary}

The combined single-dish and interferometric observations of L1689N
presented here show clear evidence of an interaction between the
prestellar core seen in the dust continuum and line emission of
deuterated molecular tracers, notably \nhhd, \nddd, \nndp, and DNC,
and the molecular outflow emanating from the nearby solar-type
protostar IRAS~16293. \emph{Herschel} observations of water and singly
deuterated ammonia show that the outflow wraps around and largely
avoids the prestellar core. A shift in the \nhhd\ line velocity and
width is seen across the core, with the narrowest line profiles
observed toward the north-eastern part of the core. We suggest that
the shock associated with the outflow has already propagated through
this part of the core and we see quiescent, cold, shock-compressed,
dense gas, blueshifted with respect to the systemic velocity of the
cloud. The western/south-western part of the core is still interacting
with the outflow, as evidenced by the broader, more disturbed line
profiles.

The \nndp\ emission observed with the ALMA Compact Array shows an
elongated distribution, extending south/south-west from the
single-dish \nhhd\ peak. \nddd\ shows a similar morphology, although
the peak of the emission is shifted \about 2.5\arcs\ east compared to
\nndp, possibly due to optical depth effects. The 970~\mic\ dust
continuum emission observed with the ACA is offset \about 5\arcs (one
ACA synthesized beam) to the south with respect to the \nndp\ and
\nddd\ emission peaks. A possible explanation is that the material in
the northern part of the core is largely uniformly distributed and has
not yet fragmented into compact sources that can lead to a second
generation of star formation in the region. The continuum emission is
thus largely resolved out by the interferometer. However, a compact
dust source with a size of \about 1100~au and an \hh\ mass of \about
0.2--0.4~M\solar\ is present in the southern part of the core. The
sensitivity of the current ACA \nddd\ data is not sufficient to
determine whether this source is collapsing and the high-SNR \nndp\
spectra are affected by the complex HFS.

The observations presented here provide clear evidence of a physical
interaction between the prestellar core in L1689N and the blue lobe of
one of the outflows driven by IRAS16293. How frequent such
interactions may be in nearby low-mass star forming regions is
difficult to answer quantitatively. \cite{hiramatsu10} suggested that
a similar interaction may occur in Barnard 1. They concluded that that
outflows cannot sustain the turbulence in the main Barnard ~1 core,
but they are energetic enough to compensate the turbulent energy
dissipation in the neighboring, more evolved star forming region
NGC1333. Recent observations of Serpens South and NGC1333
\citep{plunkett15} also suggest that outflows are energetically
important and provide sufficient energy to sustain turbulence early in
the protocluster formation process, without disrupting the entire
cluster. Observations of low-mass star forming regions often show
bright \dcop\ emission, assumed to trace cold, dense gas, coinciding
with lobes of high-velocity CO emission rather than dust continuum
peaks, marking the locations of embedded protostars (e.g.,
\citealt{lis04}). Most of these \dcop\ sources have not been studied
in enough detail to determine whether they are true interaction
regions, or whether they in fact harbor prestellar cores, like is the
case in L1689N. \cite{roueff05} also pointed out that \nhhd\ and
\nddd\ emission often does not peak at the location of embedded
protostars, but instead at offset positions, where outflow
interactions may occur. In their sample of 6 sources where \nddd\
emission has been detected, L1689N is the only region where a physical
interaction has been clearly demonstrated, but an outflow interaction
has also been suggested in Barnard~1 \citep{hiramatsu10}. Another
candidate source is located in NGC1333, at the edge of the blue lobe
of the powerful molecular outflow associated with the HH 7--11 jet
\citep{lis04}. However, this source has not yet been studied
sufficiently to make a convincing case.

In conclusion, the observations presented here demonstrate the utility
of the fundamental rotational transitions of deuterated ammonia as an
additional tracer of the deeply embedded, pre-stellar phase of star
formation. Collisional cross-sections of \nhhd, \nddh, and \nddd\ with
\hh\ are now available \citep{daniel14, daniel16b}, allowing detailed
modeling using state-of-the art radiative transfer codes. The lines
are accessible to the current ground-based submillimeter facilities,
in particular ALMA, offering new insights into the early phases of the
star formation process.

\acknowledgments HIFI has been designed and built by a consortium of
institutes and university departments from across Europe, Canada and
the United States (NASA) under the leadership of SRON, Netherlands
Institute for Space Research, Groningen, The Netherlands, and with
major contributions from Germany, France and the US. Support for this
work was provided by NASA (\emph{Herschel} OT funding) through an
award issued by JPL/Caltech. This paper makes use of the following
ALMA data: ADS/JAO.ALMA\#2012.1.00178.S. ALMA is a partnership of ESO
(representing its member states), NSF (USA) and NINS (Japan), together
with NRC (Canada), NSC and ASIAA (Taiwan), and KASI (Republic of
Korea), in cooperation with the Republic of Chile. The Joint ALMA
Observatory is operated by ESO, AUI/NRAO and NAOJ. This work is based
upon observations with the the Caltech Submillimeter Observatory,
operated by the California Institute of Technology. C.M.W. would like
to acknowledge financial support from the Science Foundation Ireland
(Grant 13/ERC/I12907). We thank an anonymous referee for helpful
comments.

\end{document}